\documentclass[twocolumn]{aastex61}

\newcommand\aastex{AAS\TeX}

\newcommand{\rone}{0.599}
\newcommand{\rtwo}{0.625}
\newcommand{\mone}{0.661}
\newcommand{\mtwo}{0.608}

 \accepted{July 5, 2019}

\shorttitle{\aastex\ Magnetic Inflation and Stellar Mass III}
\shortauthors{Healy et al.}

\begin{document}

\title{Magnetic Inflation and Stellar Mass III: Revised Parameters for the Component Stars of NSVS 07394765}

\correspondingauthor{Brian F. Healy}
\email{bfhealy@jhu.edu}

\author[0000-0002-7718-7884]{Brian F. Healy}
\affil{Department of Physics and Astronomy, Johns Hopkins University, 3400 North Charles Street, Baltimore, MD 21218, USA}
\affil{Department of Astronomy \& Institute for Astrophysical Research, Boston University, 725 Commonwealth Avenue, Boston, MA 02215, USA}

\author[0000-0001-9797-0019]{Eunkyu Han}
\affil{Department of Astronomy \& Institute for Astrophysical Research, Boston University, 725 Commonwealth Avenue, Boston, MA 02215, USA}

\author[0000-0002-0638-8822]{Philip S. Muirhead}
\affil{Department of Astronomy \& Institute for Astrophysical Research, Boston University, 725 Commonwealth Avenue, Boston, MA 02215, USA}

\author{Brian Skiff}
\affiliation{Lowell Observatory, 1400 West Mars Hill Road, Flagstaff, AZ 86001, USA}

\author{Tom Polakis}
\affiliation{Command Module Observatory, 121 West Alameda Drive, Tempe, AZ 85282, USA}

\author[0000-0002-3091-8061]{Anneliese Rilinger}
\affil{Department of Astronomy \& Institute for Astrophysical Research, Boston University, 725 Commonwealth Avenue, Boston, MA 02215, USA}

\author[0000-0002-9486-818X]{Jonathan J. Swift}
\affiliation{The Thacher School, 5025 Thacher Road, Ojai, CA 93023, USA}

\begin{abstract}
We perform a new analysis of the M dwarf-M dwarf eclipsing binary system NSVS 07394765 in order to investigate the reported hyper-inflated radius of one of the component stars. Our analysis is based on archival photometry from the Wide Angle Search for Planets (WASP), new photometry from the 32 cm {Command Module Observatory (CMO) telescope in Arizona and the 70 cm telescope at Thacher Observatory in California}, and new high-resolution infrared spectra obtained with the Immersion Grating Infrared Spectrograph (IGRINS) on the Discovery Channel Telescope. The masses and radii we measure for each component star disagree with previously reported measurements. We show that both stars are early M-type main-sequence stars without evidence for youth or hyper-inflation ($M_1= \mone\ ^{+0.008}_{-0.036}\ \rm{M_\Sun}$, $M_2= \mtwo\ ^{+0.003}_{-0.028}\ \rm{M_\Sun}$, $R_1= \rone\ ^{+0.032}_{-0.019}\ \rm{R_\Sun}$, $R_2= \rtwo\ ^{+0.012}_{-0.027}\ \rm{R_\Sun}$), and we update the orbital period and eclipse ephemerides for the system. We suggest that the likely cause of the initial hyper-inflated result is the use of moderate-resolution spectroscopy for precise radial velocity measurements.

\end{abstract}


\section{Introduction}\label{intro}

The faintest and coolest stars in the Milky Way make up for their dimness with their sheer number: over 70$\%$ of the stars in the galaxy are main-sequence M dwarf stars \citep[e.g.][]{chabrier_imf}. These stars typically have a temperature range of 2300-3800 $K$, a mass range of $\sim$0.075-0.60 $M_{\sun}$, and a radius range of 0.08-0.62 $R_{\sun}$. Since the maximum luminosity of an M dwarf is less than 10$\%$ that of the sun, it can be difficult to investigate their properties, especially when they are isolated. When their presence is observed within a detached, non-interacting eclipsing binary (EB), however, the opportunity to learn about both stars in the system greatly increases. 

Previous studies of M dwarfs in detached EBs have shown an empirical relationship between the mass and radius of these stars \citep{Torres2010}. The empirical relationship appears to show larger radii for a given mass than predictions from evolutionary models \citep[e.g.][]{Feiden2013}, and the individual M dwarfs show significant significant scatter around this relationship \citep[][]{Parsons2018}. One potential explanation for the ``inflated'' M dwarfs involves effects from magnetic fields.  In magnetically active stars, strong magnetic fields may disrupt stellar convection cells that transport energy toward the surface. The effect can be simulated using mixing length theory by increasing the mixing length parameter in stellar evolutionary models \citep{Chabrier2007}.

In this scenario, the lowered convective efficiency leads to a steeper temperature gradient, resulting in a lower stellar effective temperature $T_{\rm{eff}}$. Since the star's nuclear reaction rate and corresponding luminosity is nearly unchanged, a lower $T_{\rm{eff}}$ leads to a higher ``inflated'' radius compared to a star with higher convective efficiency. Recent work by \citet[][]{MacDonald2017} suggests that the observed inflation can be caused by magnetic fields less than 10 kG.  However, even in the most magnetically active M dwarfs, this process is not expected to inflate stellar radii beyond around $25\%$ of their non-inflated counterparts, and only for M dwarfs that are partially convective.  For fully convective M dwarfs, a different mechanism involving star spots may cause inflation through flux suppression \citep[][]{Chabrier2007}.  In a recent paper in this paper series, \citet[][]{Kesseli2018} showed that single, rapidly-rotating, fully-convective stars also appear larger than evolutionary models predict, providing evidence for flux-suppression by magnetized starspots.

Challenging these proposed scenarios are several main-sequence EBs that appear to be {\it hyper}-inflated, with radii far greater than either the empirical trend or model predictions for their masses, even after considering the effects of strong magnetic fields.   One example is T-Cyg1-12664, a main-sequence low-mass EB with {\it Kepler} photometry, that initially appeared to contain hyper-inflated components \citep[][]{Cakirli2013b,Iglesias2017}.  In a previous paper as part of this series, \citet{Han2017} showed that neither component star of T-Cyg1-12664 showed evidence of hyper-inflation, and that both are consistent with the empirical mass-radius trend seen in typical EBs.  They attributed the discrepancy to the use of high-resolution near-infrared spectroscopy, which provided higher-fidelity radial velocity observations of both component stars.

\begin{figure*}
    \centering
    \includegraphics[scale=0.5]{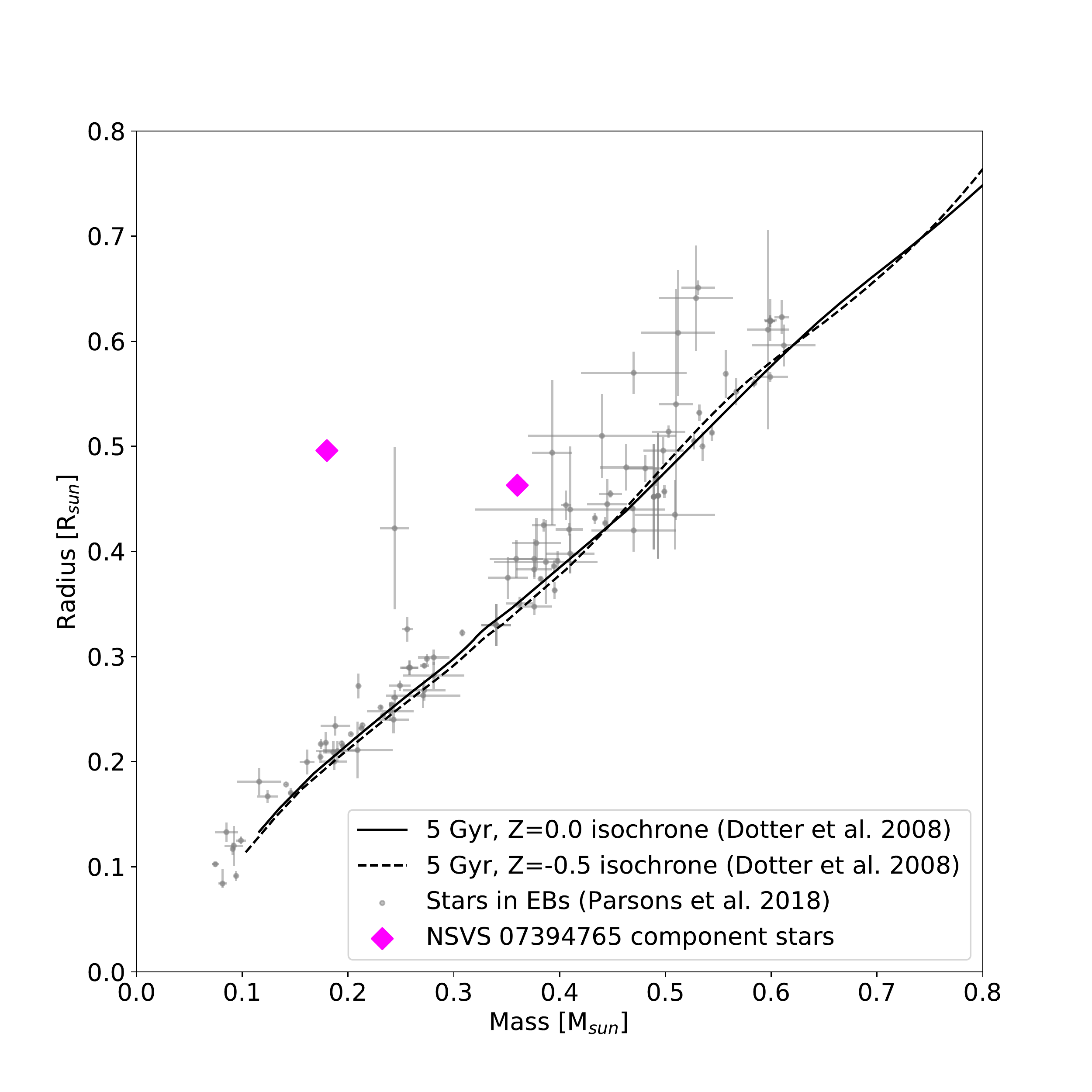}
    \caption{Mass-radius plot of Dartmouth 5 Gyr stellar isochrones \citep{dartmouthiso} for metallicities of 0.0 and -0.5 dex and stars in detached eclipsing binaries with reliable measurements from the literature \citep[see][Table A1]{Parsons2018}. The reported measurements of the components of NSVS 0739 are shown in magenta, with reported uncertainties that are smaller than the size of the points. There is a greater than 20$\sigma$ discrepancy between the models and the reported masses and radii.}
    \label{fig:dartmouth}
\end{figure*}

Another hyper-inflated detached EB, NSVS 07394765 (hereafter NSVS 0739), is reported to have an M dwarf component with a radius and mass of 0.50 $R_{\sun}$ and 0.18 $M_{\sun}$, respectively \citep{cakirli1}. This radius is more than twice the predictions from either stellar evolutionary models \citep{dartmouthiso} or empirical trends (Figure \ref{fig:dartmouth}), deviating significantly even from predictions involving magnetic inflation. The other M dwarf in this system is reported to be less inflated, with parameters 0.46 $R_{\sun}$ and 0.36 $M_{\sun}$. 

These parameters suggest that NSVS 0739 is an ideal system for testing theories of stellar inflation. For an M dwarf to be this inflated, it must be either a nascent star that is in the process of contracting (pre-main sequence), or the result of some unknown mechanism. If it is young, the star would offer valuable information about the evolutionary track of M dwarfs \citep[e.g.][]{kraus2015, Gillen2017}.

Thus, we investigated NSVS 0739 to determine if one of the components is in fact a pre-main sequence star.  In our examination of this system, we found that neither M dwarf component of NSVS 0739 is hyper-inflated or even moderately inflated compared to the mass-radius trend (see Section 3.2). Instead, the revised parameters are in statistical compatibility with the empirical mass-radius relation. We argue, similar to \citet{Han2017}, that our use of high-resolution infrared spectroscopy to measure radial velocities improved the accuracy of those measurements. 

In Section \ref{sec:obs} of this paper, we describe our photometric and spectroscopic data and the reduction of these data. In Section \ref{sec:results}, we present the results from fitting an eclipsing binary model to the data, using the same procedure outlined in \citet{Han2017}. Section \ref{sec:discussion} discusses discrepancies between our results and published values, and Section \ref{sec:conclusion} states our conclusions from this work.

\section{Data and Reduction}
\label{sec:obs}

\subsection{Archival Photometry}

\begin{figure*}
    \centering
    \includegraphics[scale=0.65]{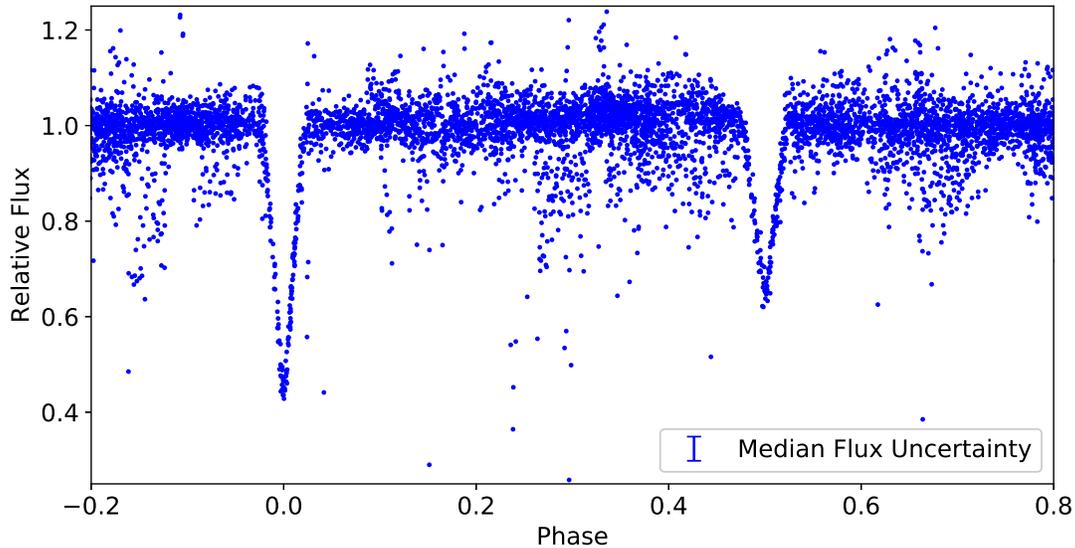}
    \caption{Phase-folded WASP light curve for NSVS 0739. The data have been cleaned of erroneous in-eclipse data points. {The legend shows the median flux uncertainty.}}
    \label{fig:nsvs_full}
\end{figure*}

We accessed the NASA Exoplanet Archive \citep{exoplanet} to download publicly-available data for NSVS 0739 from the Wide-Angle Search for Planets \citep[WASP,][]{Butters2010}. {The WASP passband ranges from $\sim$400-700 nm, roughly encompassing the Sloan {\it g} and {\it r} bands}. The observations were made between Sep. 2004 and Apr. 2008 for a total of 6718 photometric data points. With a $V$-band magnitude of 13.0, NSVS 0739 is near the limiting magnitude of the survey, introducing noticeable noise into the light curve. Nonetheless, Figure \ref{fig:nsvs_full} shows that WASP clearly detected both eclipses for the system. See Table 1 for the details of the WASP observations.

We removed two nights of WASP data for which there were data points significantly deviating from the expected in-eclipse value despite having a phase corresponding to an eclipse. These points were likely caused by adverse weather conditions during these nights. We also established a maximum relative flux limit of 1.1 to exclude extreme increases in flux corresponding to weather or flares on the component stars. {We converted time units from the archive-supplied HJD to $\rm{BJD}_{\rm TDB}$ using a calculator by \citet{eastman2010}.}

\subsection{New Photometry}

\begin{figure*}
    \centering
    \includegraphics[scale=0.55]{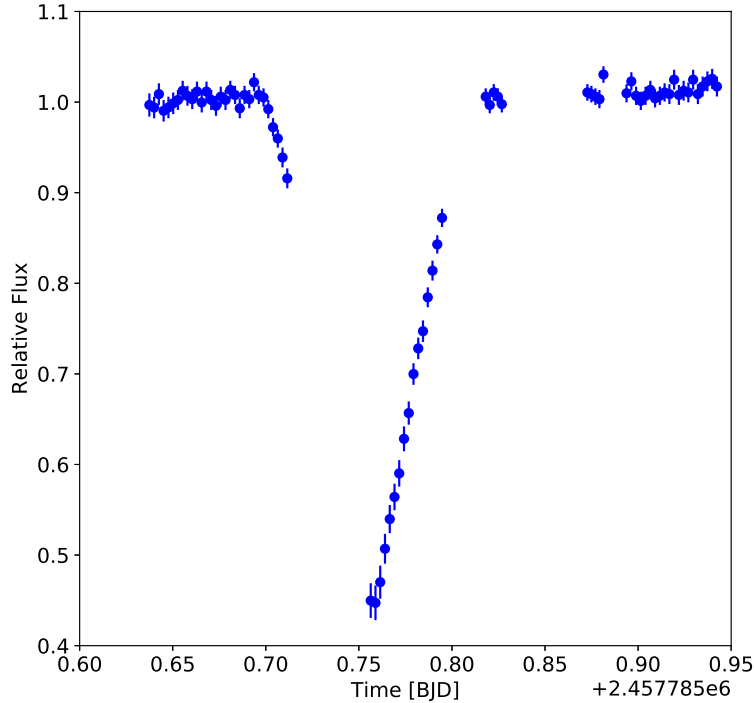}
    \caption{Additional $V$-band primary eclipse light curve observed at Command Module Observatory on UT Feb. 1, 2017.}
    \label{fig:newpri}
\end{figure*}

We also obtained a new NSVS 0739 primary eclipse observation on UT Feb. 1, 2017 using the 32 cm Dall-Kirkham telescope at Command Module Observatory in Tempe, AZ. The detector is a thermo-electrically cooled SBIG ST-6303On CCD.  The night was photometric, and we acquired 180-second exposures in the Johnson $V$ band. {We used the commercial software package MPO Canopus to perform aperture photometry on NSVS 0739 and 4 reference stars with 15-arcsec apertures and sky annulus subtraction. This software specializes in asteroid and variable star analysis, offering a graphical interface for image calibration, astrometry, and photometry.} We used the AAVSO Photometric All-Sky Survey (APASS) DR9 to supply $V$ magnitudes for the reference stars. Because we  observed multiple targets on this night, there are gaps in coverage of the NSVS 0739 eclipse (Figure \ref{fig:newpri}).

\begin{figure*}
    \centering
    \includegraphics[scale=0.55]{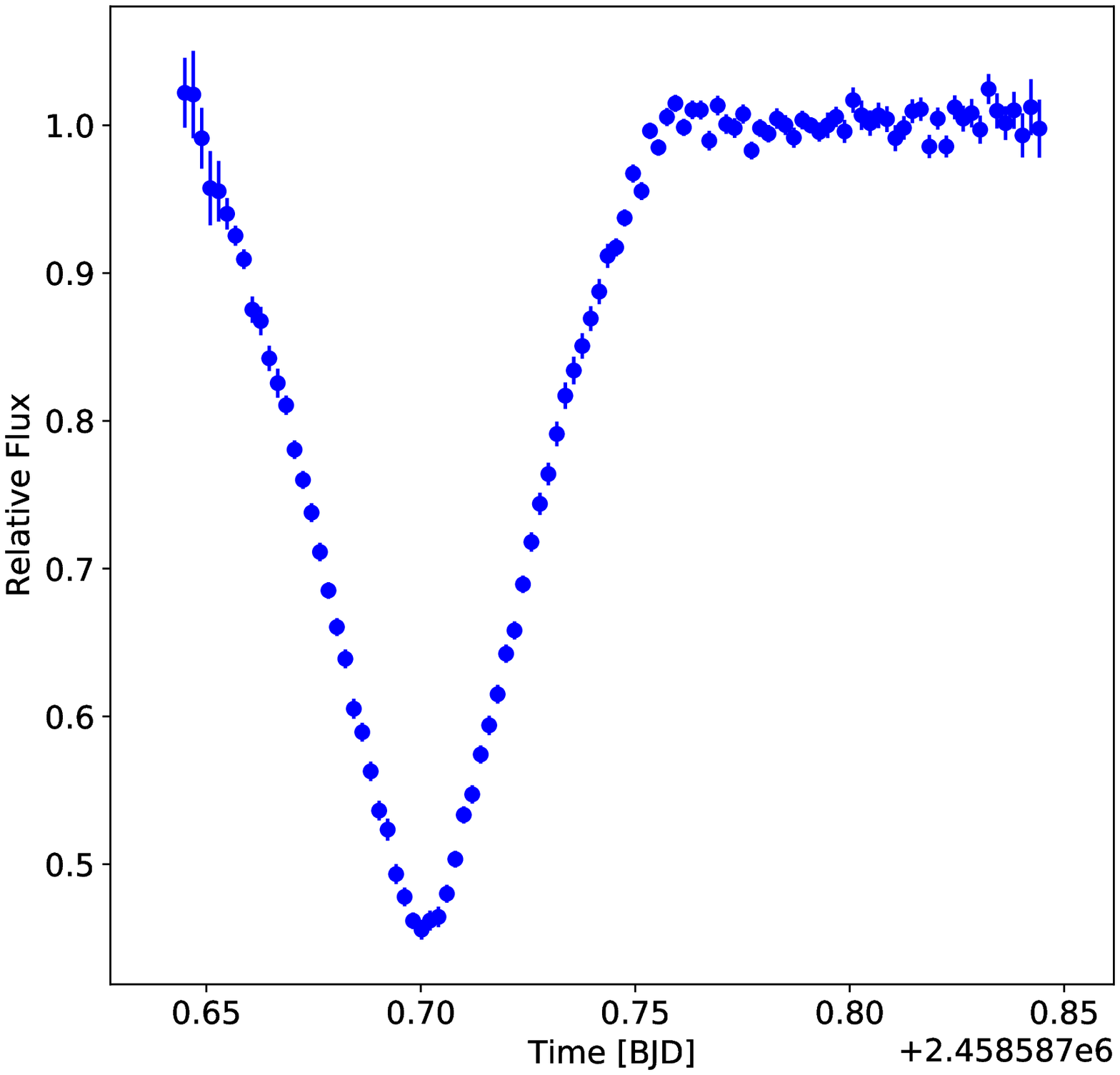}
    \caption{Additional $V$-band primary eclipse light curve observed at Thacher Observatory on UT Apr. 14, 2019.}
    \label{fig:newpri_thacherv}
\end{figure*}

{We observed another primary eclipse (Figure \ref{fig:newpri_thacherv}) on UT Apr. 14, 2019 with the 0.7 m telescope at Thacher Observatory in Ojai, CA \citep{thacher1,thacher2}. The new observation was made in the Johnson $V$ band with integration times of 1 minute. We reduced the data using the \texttt{astropy} utilities in Python \citep{astropy1,astropy2}. We performed aperture photometry on NSVS 0739 and two nearby reference stars that we tested for stability and high SNR. Given the variability of the seeing, we optimized the aperture used in each image to maximize SNR on NSVS 0739, and we chose sky radii from stacked images to avoid background sources in areas outside the wings of the PSF. The final light curve was stable and did not require us to fit the out-of-eclipse data for a trend in airmass or time.}

\begin{table}[]
    \centering
    \caption{Description of WASP observations for NSVS 0739.}
    \begin{tabular}{c|c}
        \hline
        \hline
        Coordinates (RA Dec) & {8$^{\rm{h}}$ 25$^{\rm{m}}$ 51$^{\rm{s}}$.894, 24$^{\circ}$ 27' 4.60''} \\
        WASP Magnitude & 13.17819 \\
        Start time (BJD) & 2453261.742889 \\
        End time (BJD) & 2454575.437458 \\
        Number of points & 6718 \\
        \hline
    \end{tabular}
    \label{table:photometric}
\end{table}

\subsection{Spectroscopy}

\begin{figure*}
    \centering
    \includegraphics[scale=0.6]{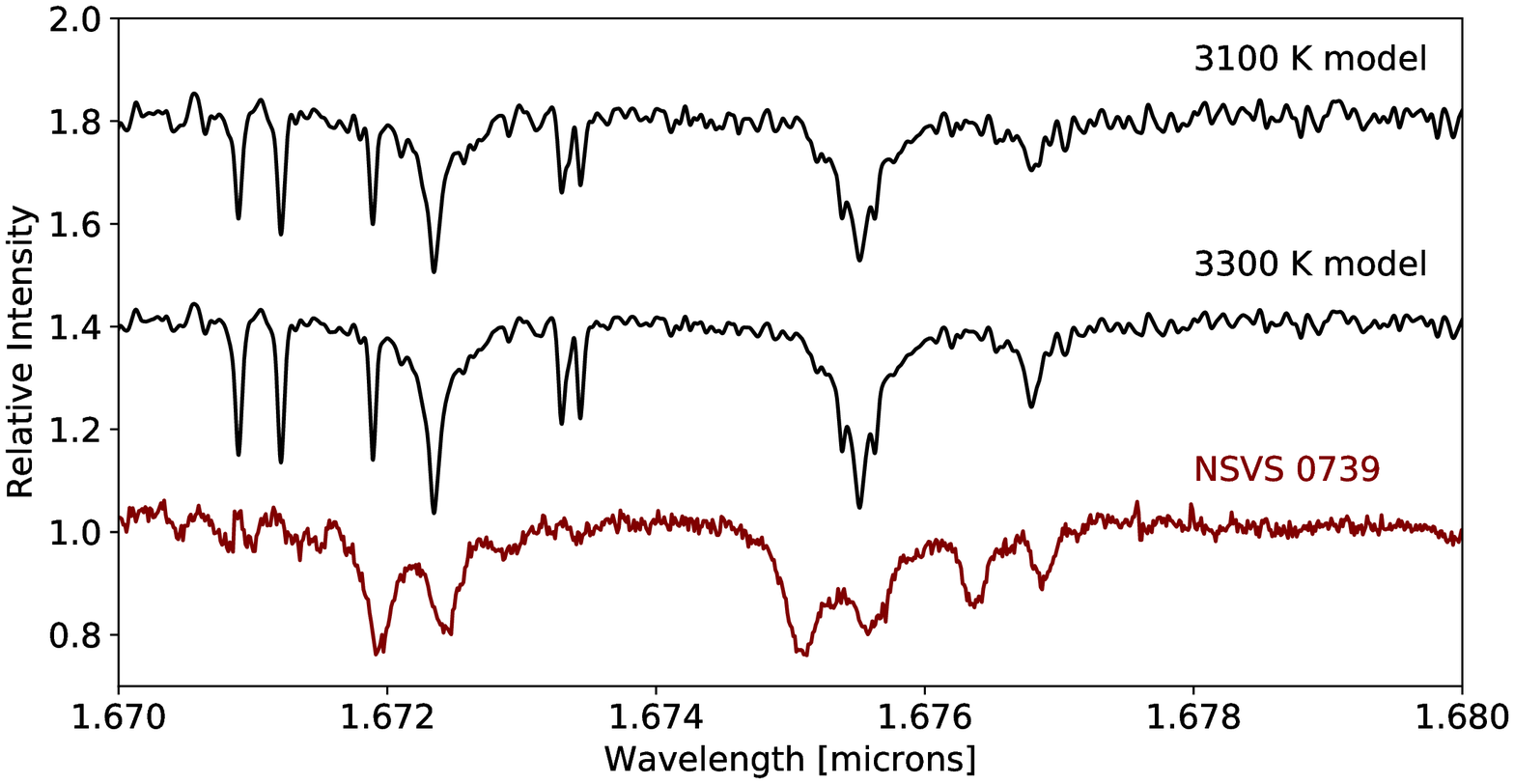}
    \caption{Example IGRINS 10th-order $H$-band spectrum for NSVS 0739 (maroon), observed on UT Nov. 10, 2016. The double-lined nature of this binary system is apparent in the duplication of an Al I triplet in this order \citep{cushing2005}. Two BT-Settl model spectra, representing stars with 3100 K and 3300 K effective temperatures, are plotted above with a vertical offset.}
    \label{fig:igrins}
\end{figure*}

We used the Immersion Grating Infrared Spectrograph \citep[IGRINS,][]{igrins,igrinsdct} at the Discovery Channel Telescope to obtain spectra for NSVS 0739 at five different times. IGRINS is a cross-dispersed, near-infrared, high-resolution spectrometer covering wavelengths between 1.45 and 2.45 $\mu$m ($H$ and $K$ bands) at R $\sim$ 45,000. Calculated exposure times were intended to provide a signal-to-noise ratio of at least 10. See Figure \ref{fig:igrins} for an example $H$-band spectrum. Given that the orbital period of NSVS 0739 was on the order of two days, useful observations only needed to be separated by a few hours.

We reduced the IGRINS data following the procedures described in \citet{Han2017}.  Briefly, the first step involved feeding the raw data through the IGRINS pipeline \citep{sim2014}. We then used observations of a nearby A0 star to calibrate for telluric lines. {We observed the A0 star on the same night as each target observation, under similar weather conditions.} The telluric correction was done using the \texttt{xtellcor\_general} data reduction software package {written in IDL} (\citealp{xtellcor}). {The software propagates uncertainties in the telluric correction to the final uncertainties in the spectra.}

\begin{figure*}
    \centering
    \includegraphics[scale=0.8]{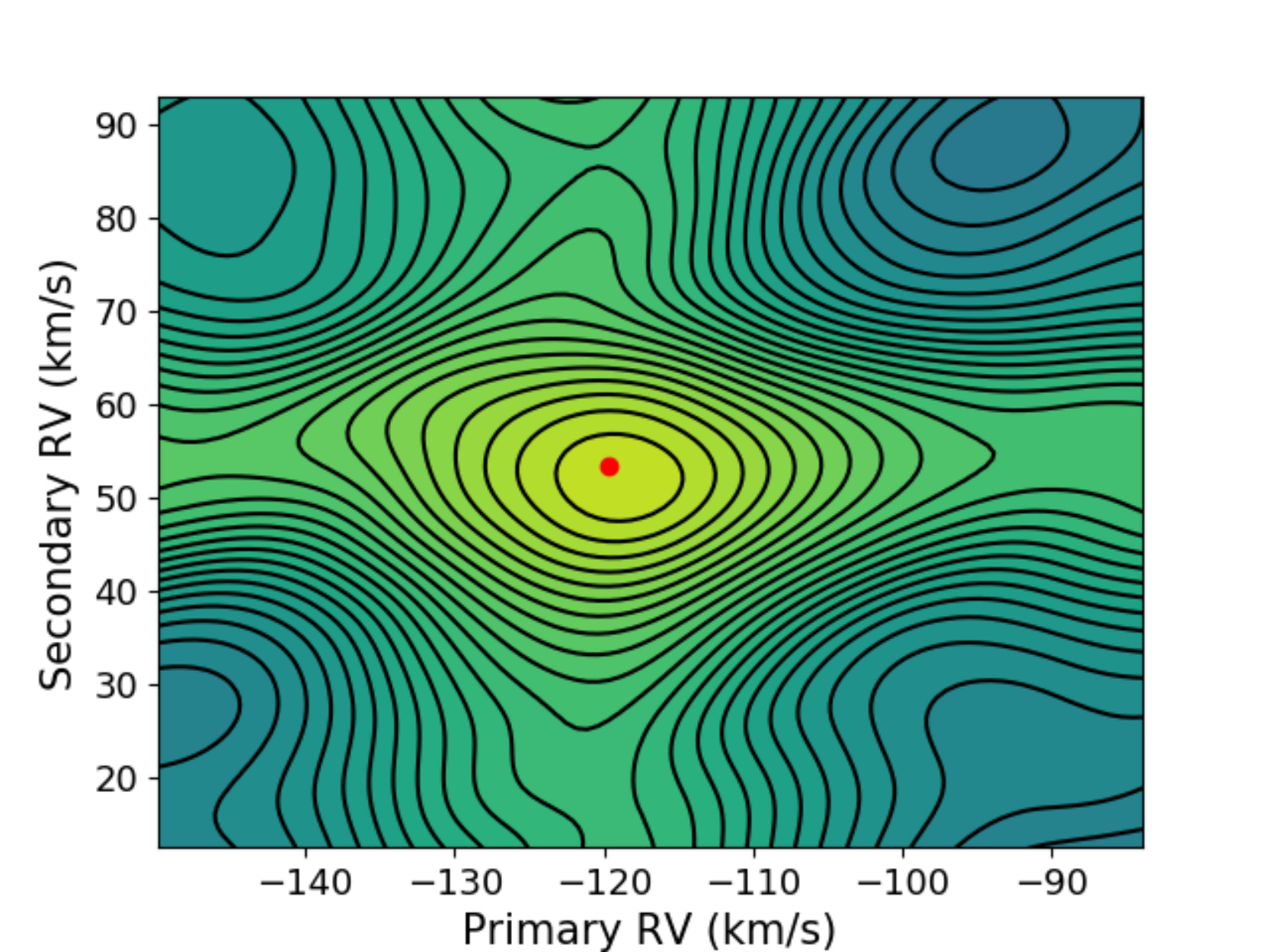}
    \caption{Example two-dimensional cross-correlation of template and target spectra {with \texttt{TODCOR} \citep{zucker1994}} to obtain primary and secondary radial velocities.}
    \label{fig:ccf1}
\end{figure*}

{Using \texttt{TODCOR} \citep{zucker1994}}, we performed a two-dimensional cross-correlation between {two template BT-Settl model spectra \citep{baraffe2015,allard2012} for stars with 3100 K and 3300 K effective temperatures (Figure \ref{fig:igrins})} and high-SNR spectral orders 8-14 of our $H$-band IGRINS data to find the radial velocity of both stars in the system (e.g.\ Figure \ref{fig:ccf1}). {Orders near the edges of the detector experience distortions that diminish the quality of derived RVs, and other unused orders contain large telluric features that are not sufficiently corrected by the A0 calibration spectrum.} We obtained uncertainties on the RVs by performing cross-correlation with each order independently and taking a standard deviation of the mean of the results. We did not use the lower-SNR $K$-band data from IGRINS because the cross-correlation functions were not as definitive in this band. To account for the motion of the Earth around the Sun, we performed a barycentric correction using the tools of \citet{barycorr}. Table 2 lists the five new radial velocity points.

\section{Analysis and Results}
\label{sec:results}

\begin{figure*}
    \centering
    \includegraphics[scale=0.4]{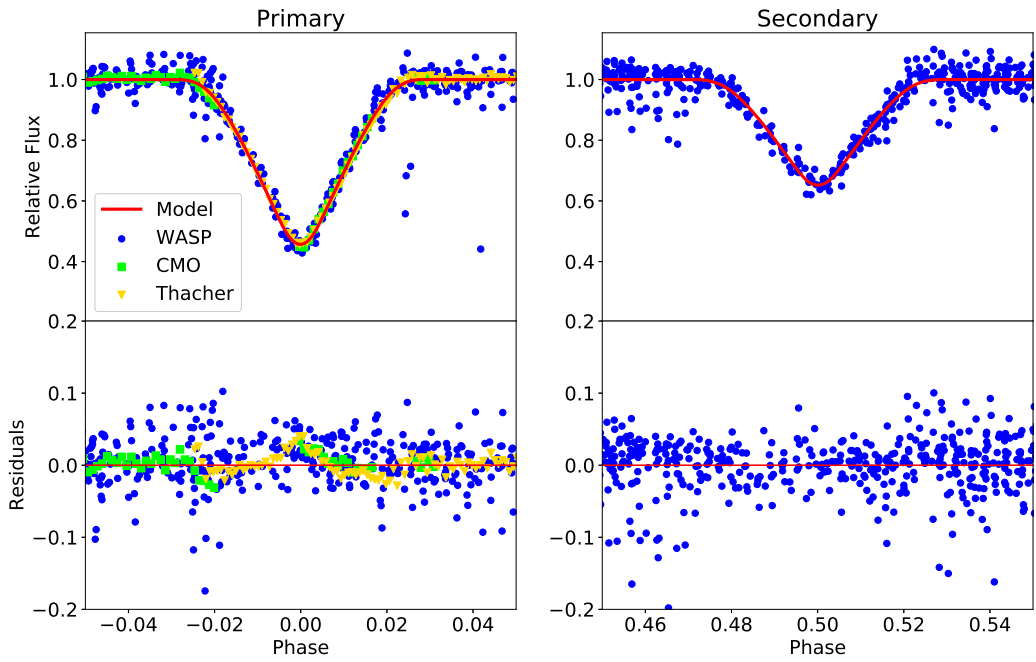}
    \caption{Model fit and residuals for WASP, CMO and Thacher primary and secondary eclipses. {The shape of the CMO and Thacher residuals illustrates the difference between the WASP and $V$ passbands.}}
    \label{fig:mainfig}
\end{figure*}

\begin{deluxetable*}{lcccccrrccccrcc}
\tabletypesize{\large}
\tablewidth{0pt}
\tablecolumns{5}
\tablecaption{New barycenter-corrected NSVS 0739 radial velocities calculated from IGRINS $H$-band spectra.}
\label{table:newrv}
\tablehead{\colhead{Band}&
          \colhead{Time (BJD$_{\rm{TDB}}$)}&
          \colhead{Phase}&
          \colhead{Primary RV (km/s)}&
		  \colhead{Secondary RV (km/s)}&
		  \cr
		  \colhead{(1)}&
		  \colhead{(2)}&
		  \colhead{(3)}&
		  \colhead{(4)}&
		  \colhead{(5)}}
		  
\startdata
$H$ & 2457702.86978 & 0.411 & $-48.2 \pm 2.5$  & $43.2 \pm 2.5$ \\
$H$ & 2457702.94347 & 0.444 & $-31.4 \pm 2.6$ & $27.8 \pm 1.2$ \\
$H$ & 2458022.00737 & 0.287 & $-84.8 \pm 1.2 $ & $84.8 \pm 4.8$ \\
$H$ & 2458023.02147 & 0.735 & $80.1 \pm 1.1 $ & $-94.3 \pm 3.6$ \\
{$H$} & {2458473.95167} & {0.788} &  ${76.9 \pm 6.4}$ & ${-94.5 \pm 11.2}$ \\
\enddata
\end{deluxetable*}

\begin{figure*}
    \centering
    \includegraphics[scale=0.5]{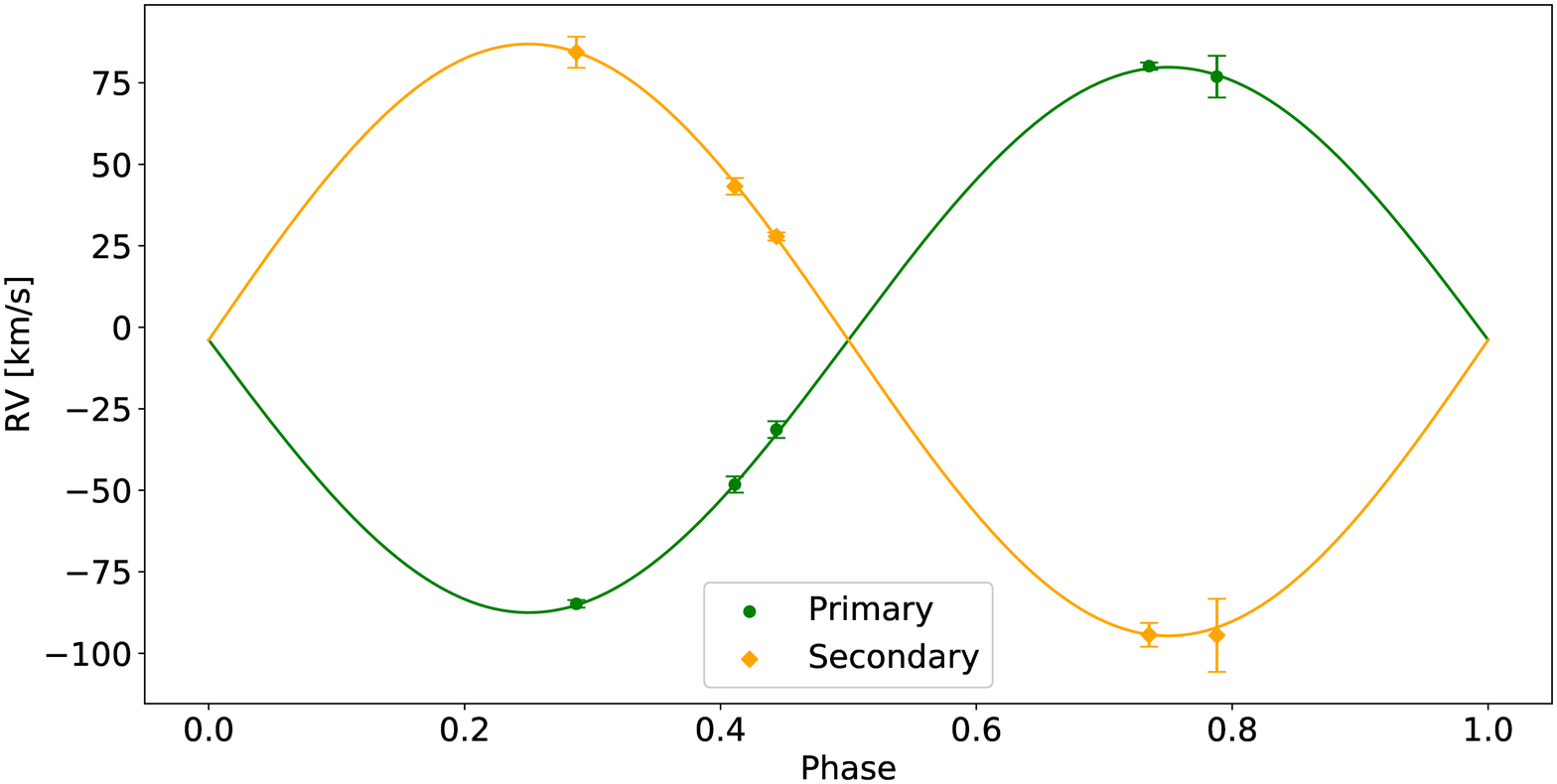}
    \caption{Radial velocity fit to the five IGRINS data points.}
    \label{fig:newrvs}
\end{figure*}

\subsection{Model Fitting}
 {Under the assumption that the passbands of WASP, CMO and Thacher data overlapped,} we fit a model to the photometry and RVs based on the \texttt{eb} software by \citet{Irwin2011}. {This code generates simulated photometry and radial velocity curves. The upper rows of Table 3 show the model parameters that were fitted. Under the assumption of no third light, and neglecting the effect of light-travel time due to the system's nearly equal-mass components and low eccentricity,} we made an initial least-squares fit using \texttt{mpfit} \citep{Markwardt2009}. Initiating uniform priors for each parameter centered on our results from \texttt{mpfit}, we performed a Markov Chain Monte Carlo (MCMC) exploration of parameter space using \texttt{emcee} \citep{Foreman-Mackey2013}. {For the MCMC, we established a normal likelihood function (LF) of the form 
 \begin{equation}
     \ln(\rm{LF}) = -0.5 * \sum_{i=1}^{n_{points}} \bigg[ \frac{(y_{i} - \hat{y}_{i})^2}{\sigma_{i}^2} +  \ \ln(2 \pi \sigma_{i}^2) \bigg],
 \end{equation}
 summing over all points (n$_{\rm{points}}$) in the phase-folded data ($y_{i}$) in comparison with model points ($\hat{y}_{i}$) generated by \texttt{eb} and the uncertainty of each data point, $\sigma_{i}$. For each step out of the total 50,000 and each chain out of 100, we performed Affine-Invariant sampling of the fitted parameters with the \texttt{EnsembleSampler} class of \texttt{emcee}. After visual inspection of the chains for each parameter, we discarded a burn-in of the first 10,000 steps to prevent our results from being biased towards the prior values.} 
 
 {Due to the small number of RV data points compared to the plentiful photometric points, we performed the latter fit separately from the former, with the same number of steps, chains and burn-in. This separation of fits ensured that the overall determination of goodness-of-fit was not dominated only by photometry (see Section 3.2 of \citet{Han2017}, who also used this fitting method).} We show the final photometric fit (including WASP, CMO {and Thacher eclipses}) and residuals in Figure \ref{fig:mainfig}. {The residual structure visible in the CMO and Thacher primary eclipse fits shows the limitation of the assumption that the WASP passband overlaps with these $V$-band observations. Nonetheless, these new data helped to better constrain the period and time of mid-primary eclipse by roughly quadrupling the time baseline of eclipse observations and supplementing WASP data with higher-cadence coverage.} The primary and secondary radial velocity fit appears in Figure \ref{fig:newrvs}. 

{We fit limb darkening coefficients for each star using the square-root model demonstrated to be effective for low-mass M dwarfs in \citealp{Claret1998}. During fitting, we parametrized limb darkening in terms of $q1$ and $q2$ from \citealp{Kipping2013}. Though the final fit did not provide strong constraints on these coefficients; the sampling of a wide variety of limb darkening coefficients induced additional variation in the best-fit parameters for each step, widening the distribution of calculated radius values compared to a fit with better-constrained limb darkening coefficients. Therefore, limb darkening uncertainties are incorporated into the error of the other fitted parameters.}

\subsection{Results}
\begin{figure*}
    \centering
    \includegraphics[scale=0.375]{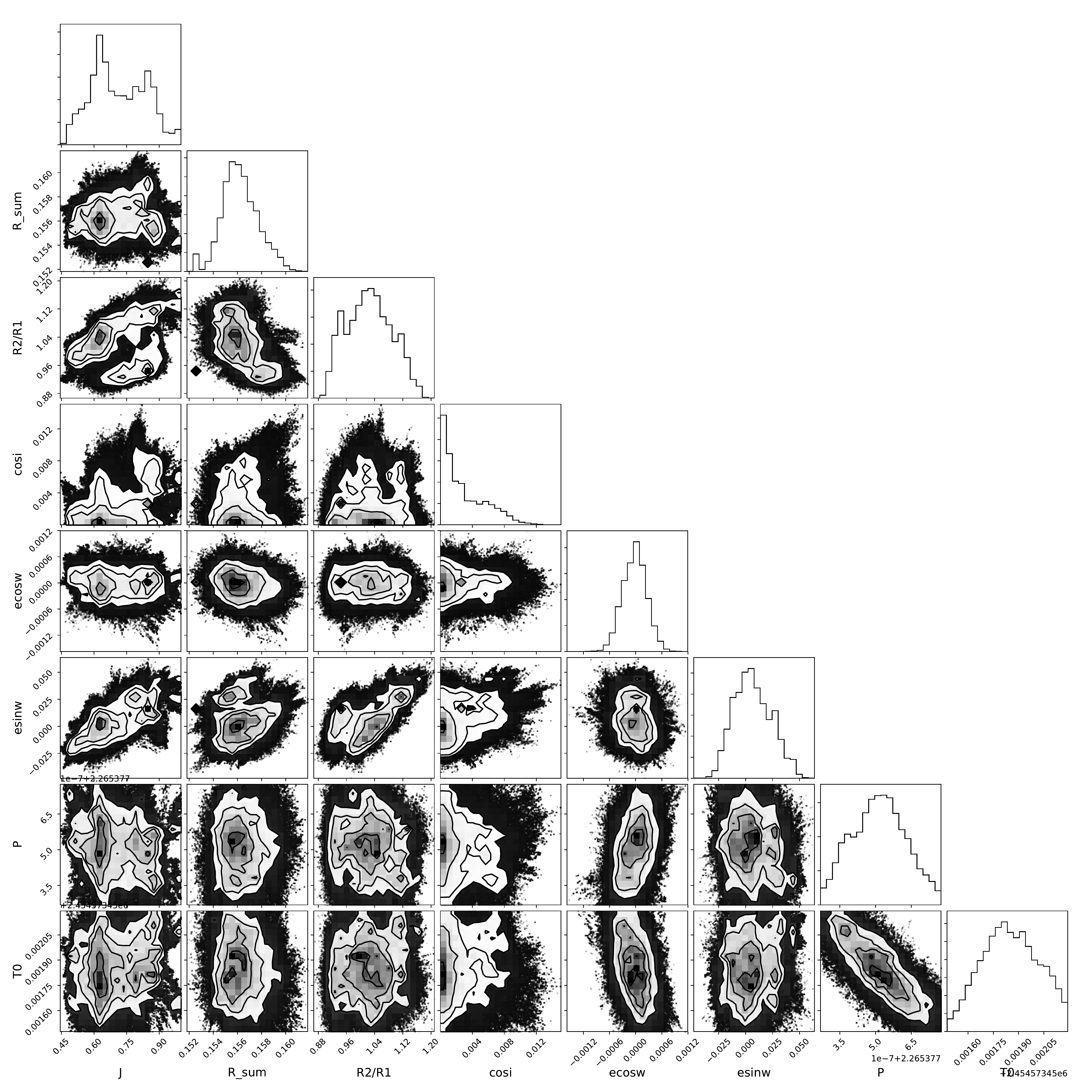}
    \caption{Triangle plot of fitted NSVS 0739 light curve parameters.}
    \label{fig:lctri}
\end{figure*}

\begin{figure*}
    \centering
    \includegraphics[scale=0.5]{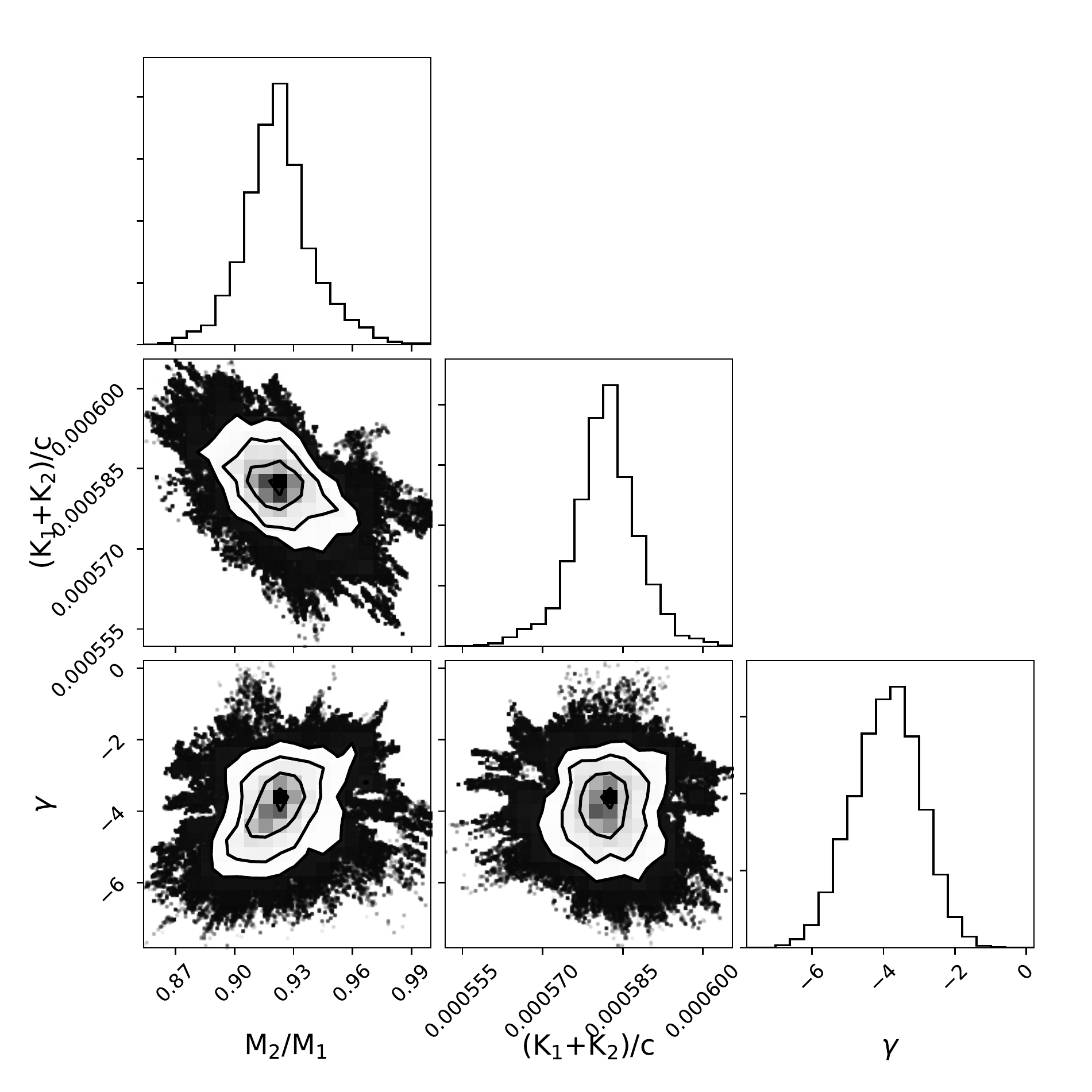}
    \caption{Triangle plot of fitted NSVS 0739 RV parameters.}
    \label{fig:rvtri}
\end{figure*}

{The MCMC run yielded $100$ chains of $40,000$ values for every fitted parameter (after discarding the burn-in steps). To solve for the desired results and their uncertainties, we calculated each final parameter from its distribution of all $4 \times 10^{6}$ values. We adopted each distribution's maximum-likelihood value to be the reported parameter value and computed its difference from the 16th and 84th percentiles of each distribution to establish the 1$\sigma$ confidence intervals reported as our uncertainties. For the eccentricity parameter, we instead used the 0th and 68th percentiles for error bars, because the nearly circular system does not have a normal distribution about the highest-likelihood value. Table 3 lists the maximum-likelihood fitted and calculated parameters with their 1$\sigma$ uncertainties. Note that to minimize the results' dependence on stellar atmospheric models, we do not compute the stellar effective temperature or luminosity ratio. See Section 4.5 of \citet{Han2017} for a further explanation. We show triangle plots \citep{dfmcorner} for the photometric and RV fits in Figures \ref{fig:lctri} and \ref{fig:rvtri}, respectively. 
We note that our ephemeris predicts future eclipses at significantly different times than the discovery paper's.} 

\begin{figure*}
    \centering
    \includegraphics[scale=0.7]{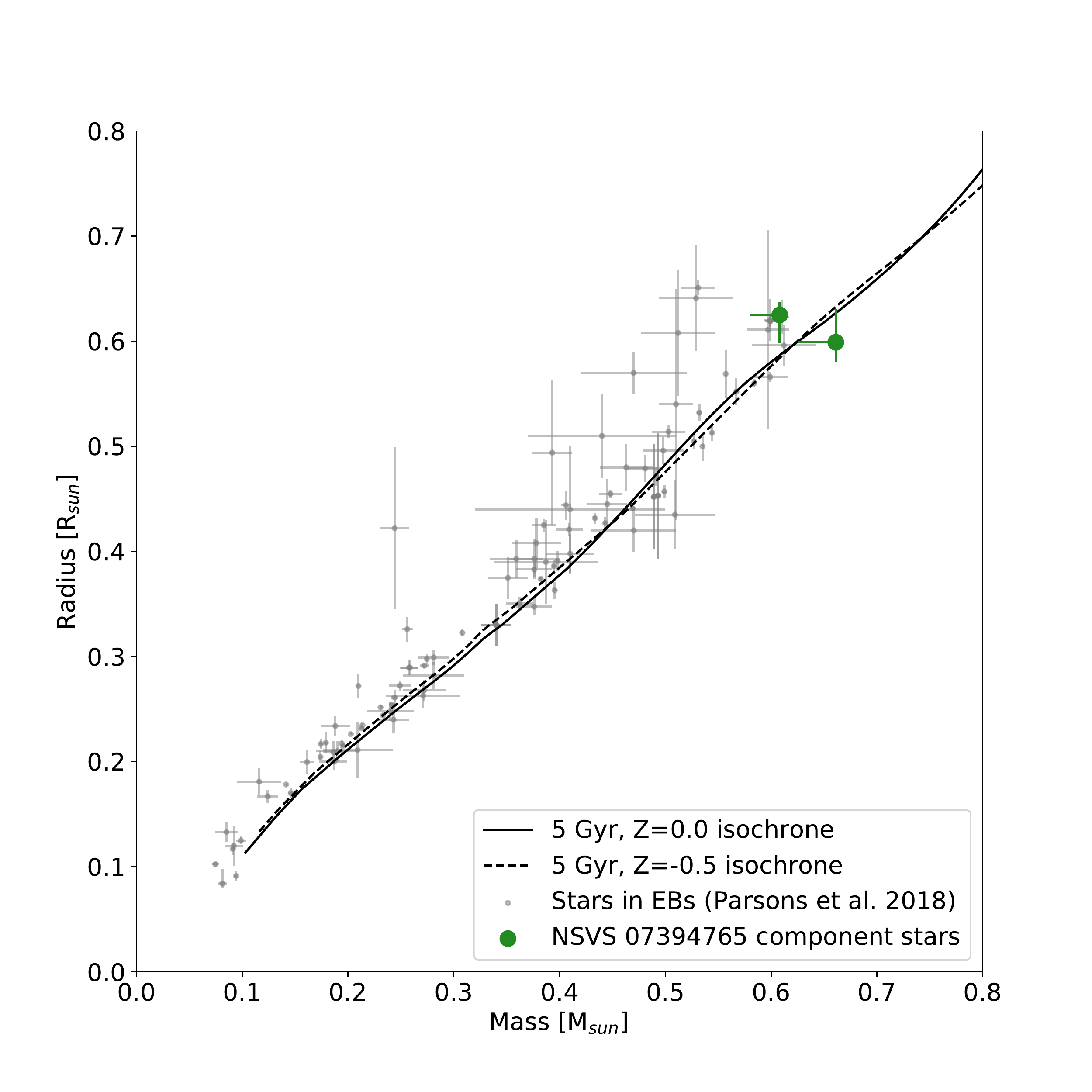}
    \caption{Mass-radius plot of theoretical 5 Gyr isochrones, stars in eclipsing binaries from \citet{Parsons2018}, and our revised parameters for NSVS 0739 and their uncertainties.}
    \label{fig:coolebs}
\end{figure*}

After our analysis, the stars now fall into statistical agreement with the empirical and theoretical mass-radius trends for M dwarfs (Figure \ref{fig:coolebs}). We discuss our confidence in the new masses and radii, along with a possible cause for the initial hyper-inflated results, in Section 4.

\begin{deluxetable*}{lcccccrrccccrcc}
\tabletypesize{\large}
\tablewidth{0pt}
\tablecolumns{3}
\tablecaption{NSVS 0739 fitted and calculated parameter descriptions, maximum-likelihood values and 1$\sigma$ uncertainties.}
\label{tbl:finalparams}
\tablehead{\colhead{Fitted Parameter}&
          \colhead{Description} &
          \colhead{NSVS 07394765}&
		  \cr
		  \colhead{(1)}&
		  \colhead{(2)} &
		  \colhead{(3)}}

\startdata
$J$ & Central surface brightness ratio &  $0.66 \ ^{+0.20}_{-0.06}$  \\
$(R_{1} + R_{2})/a$ & Fractional radii sum over semi-major axis & $0.1555 \ ^{+0.0022}_{-0.0006} $ \\
$R_{2}/R_{1}$ & Radius ratio & $1.043 \ ^{+0.053}_{-0.093}$  \\
$\cos i$ & Cosine of orbital inclination & $0.0041 \ ^{+0.0020}_{-0.0036}$ \\
$P$ & Orbital period & $2.26537743 \ ^{+0.00000021}_{-0.00000005}$ days  \\
$T_{0}$ & Time of primary mid-eclipse & $2454573.45195 \ ^{+0.00011}_{-0.00029} \ \rm{BJD_{TDB}}$ \\
$e \cos \omega$ & Eccentricity $\times$ cosine of argument of periastron & $-0.00011 \ ^{+0.00035}_{-0.00020}$ \\
$e \sin \omega$ & Eccentricity $\times$ sine of argument of periastron & $-0.001 \ ^{+0.027}_{-0.010}$ \\
$\gamma$ & Center of mass system velocity & $-3.5 \ ^{+0.5}_{-1.3}$ km/s \\
$q$ & Mass ratio ($M_{2}/M_{1}$) & $0.921 \ ^{+0.017}_{-0.014}$ \\
$K_{tot}/c$ & Sum of radial velocity semi-amplitudes / speed of light & $0.0005860 \ ^{+0.0000015}_{-0.0000096}$ \\
$u_{11}$ & Linear limb darkening coefficient, star 1 & $-0.22 \ ^{+0.60}_{-0.54}$ \\
$u_{21}$ & Square root limb darkening coefficient, star 1 & $0.43 \ ^{+0.45}_{-0.87}$ \\
$u_{12}$ & Linear limb darkening coefficient, star 2 & $0.9 \ ^{+0.7}_{-1.4}$ \\
$u_{22}$ & Square root limb darkening coefficient, star 2 & $0.1 \ ^{+1.0}_{-0.6}$ \\
\hline
Calculated Parameter & Description & NSVS 07394765 \\
\hline
$e$ & Eccentricity & $0.001\ ^{+0.017}_{-0.001}$ \\
$i$ & Orbital inclination & $89.76\ ^{+0.21}_{-0.12} \ \rm{degrees}$ \\
$a$ & Semi-major axis & $0.03651\ ^{+0.00009}_{-0.00060}$ AU \\
$K_{1}$ & Radial velocity semi-amplitude, star 1 & $84.3\ ^{+0.2}_{-1.3}$ km/s \\
$K_{2}$ & Radial velocity semi-amplitude, star 2 & $91.5\ ^{+0.7}_{-2.1}$ km/s \\
$M_{1}$ & Mass, star 1 & $\mone\ ^{+0.008}_{-0.036}$ $\rm{M_{\sun}}$  \\
$M_{2}$ & Mass, star 2 & $\mtwo\ ^{+0.003}_{-0.028}$ $\rm{M_{\sun}}$  \\
$R_{1}$ & Radius, star 1 & $\rone\ ^{+0.032}_{-0.019}$ $\rm{R_{\sun}}$  \\
$R_{2}$ & Radius, star 2 & $\rtwo\ ^{+0.012}_{-0.027}$ $\rm{R_{\sun}}$  \\
$\log g_{1}$ & Log of surface gravity, star 1 (cgs) & $4.705\ ^{+0.018}_{-0.051}$ \\
$\log g_{2}$ & Log of surface gravity, star 2 (cgs) & $4.632\ ^{+0.028}_{-0.024}$ \\
$L_{2}/L_{1}$ & Orbit-averaged photometric light ratio & $0.72\ ^{+0.31}_{-0.18}$ \\
\enddata
\label{tbl:finalparams}
\end{deluxetable*}

\section{Discussion}
\label{sec:discussion}

Our analysis supports a different argument than previous work: that 1) the NSVS 0739 M dwarfs are larger and more massive than their initial published values, 2) they are nearly equal-mass, and 3) they are not inflated. 

{Our calculated primary radial velocity semi-amplitude of $84.3 ^{+0.2}_{-1.3}$ km/s differs greatly from the previous work's value of $44 \pm 3$ km/s. This large disagreement between the two works likely accounts for the majority of the discrepancy in mass and radius results.} A possible cause for the hyper-inflated results in the previous paper may be the use of lower-resolution ($\frac{\lambda}{\Delta \lambda}$ = 7,000 compared to $\sim$ 45,000 for IGRINS) spectral observations.

{Spectral line blending can} make cross-correlation with a model spectrum more difficult, and {such an analysis} may {bias the derived} radial velocities. The dramatically different solution to the system using our high-resolution IGRINS RVs supports this hypothesis. 

\section{Conclusion}
\label{sec:conclusion}

Undertaking a new analysis of the eclipsing binary system NSVS 07394765 revealed radial velocity measurements in discrepancy with previous work. {It is likely that our high-resolution spectroscopy allowed for a more unbiased determination of RVs than was possible with the previous work's moderate-resolution observations.} The uniformity in {our} primary and secondary RV amplitudes suggests that the stellar mass ratio is near one-to-one. Moreover, these observations support a system that does not contain inflated M dwarfs.

We conclude that neither M dwarf in NSVS 07394765 is hyper-inflated. We also conclude that the system has a larger total mass that is nearly equally divided among its two components, which are likely early M dwarfs or late K dwarfs. {Our results should be considered preliminary, as the characterization of this system would still benefit from further photometric and spectroscopic observations.} This work underscores the importance of high-resolution infrared spectroscopy in the further study of low-mass stars in eclipsing binaries. As the field of astronomy moves closer to a comprehensive mass-radius-luminosity relation for M dwarfs, our results will be an important contribution to constraining these connections.

\section*{Acknowledgments}

{We thank the anonymous referee for helpful comments and critique. B.F.H. thanks {\" O}m{\" u}r {\c C}ak{\i}rl{\i} for helpful correspondence.} P.S.M. acknowledges support from the NASA Exoplanet Research Program (XRP) under Grant No. NNX15AG08G issued through the Science Mission Directorate. This research involved use of the Discovery Channel Telescope at Lowell Observatory, supported by Discovery Communications, Inc., Boston University, the University of Maryland, the University of Toledo, Northern Arizona University and Yale University.  This research involved use of the Immersion Grating Infrared Spectrometer (IGRINS) that was developed under a collaboration between the University of Texas at Austin and the Korea Astronomy and Space Science Institute (KASI) with the financial support of the US National Science Foundation under grant AST-1229522, of the University of Texas at Austin, and of the Korean GMT Project of KASI. This research has made use of the APASS database, located at the AAVSO web site. Funding for APASS has been provided by the Robert Martin Ayers Sciences Fund.

\vspace{5mm}
\facilities{DCT (IGRINS), WASP, CMO, Thacher Observatory}

\software{
\texttt{astropy} \citep{astropy1,astropy2},
\texttt{eb} \citep{Irwin2011},  
\texttt{emcee} \citep[][]{Foreman-Mackey2013},
\texttt{mpfit} \citep{Markwardt2009},
\texttt{TODCOR} \citep{zucker1994},
\texttt{xtellcor} \citep[][]{xtellcor}
 }

\bibliography{ms}
\bibliographystyle{aasjournal}

\end{document}